\documentclass[fleqn,usenatbib,usedcolumn]{mnras}

\usepackage{color,soul}

\usepackage[T1]{fontenc}

\usepackage{graphicx}	
\usepackage{amsmath}	
\usepackage{amssymb}	
\usepackage{newtxtext,newtxmath}

\title[Period Changes in LMC Cepheids]{Period Change Rates in Large Magellanic Cloud Cepheids Revisited}

\author[N. Rodriguez-Segovia et al.]{
N. Rodr\'{i}guez-Segovia,$^{1,2}$\thanks{E-mail: n.rsegovia@adfa.edu.au}
G. Hajdu,$^{3}$
M. Catelan,$^{1,4,5}$
F. Espinoza-Arancibia,$^{1,4}$
G. Boggiano,$^{1}$
\newauthor
C. Cenzano,$^{1,4}$
E. Garc\'es H.,$^{6}$
K. Joachimi,$^{1}$
C. Mu\~{n}oz-L\'{o}pez,$^{1}$
C. Ordenes-Huanca,$^{1}$
C. Orquera-Rojas$,^{1,4}$
\newauthor
P. Torres,$^{1,4}$
\'{A}. Valenzuela-Navarro$^{1,4}$
\\ 
$^{1}$Instituto de Astrof\'{i}sica, Pontificia Universidad Cat\'{o}lica de Chile, Av. Vicu\~{n}a Mackenna 4860, 7820436 Macul, Santiago, Chile\\
$^{2}$School of Science, University of New South Wales, Australian Defence Force Academy, Canberra, ACT 2600, Australia\\
$^{3}$Nicolaus Copernicus Astronomical Center, Polish Academy of Sciences, Bartycka 18, 00-716 Warsaw, Poland\\
$^{4}$Millennium Institute of Astrophysics, Nuncio Monse\~{n}or Sotero Sanz 100, Of. 104, Providencia, Santiago, Chile\\
$^{5}$Centro de Astro-Ingenier\'{i}a, Pontificia Universidad Cat\'{o}lica de Chile, Av. Vicu\~{n}a Mackenna 4860, 7820436 Macul, Santiago, Chile\\
$^{6}$Departamento de F\'isica, Facultad de Ciencias, Universidad de Chile, Casilla 653, Santiago, Chile
}

\date{Accepted XXX. Received YYY; in original form ZZZ}

\pubyear{2021}

\begin{document}
\label{firstpage}
\pagerange{\pageref{firstpage}--\pageref{lastpage}}
\maketitle

\begin{abstract}
The period-change rate (PCR) of pulsating variable stars is a useful probe of changes in their interior structure, and thus of their evolutionary stages. So far, the PCRs of Classical Cepheids in the Large Magellanic Cloud (LMC) have been explored in a limited sample of the total population of these variables. Here we use a template-based method to build observed minus computed ($O-C$) period diagrams, from which we can derive PCRs for these stars by taking advantage of the long time baseline afforded by the Digital Access to a Sky Century @ Harvard (DASCH) light curves, combined with additional data from the Optical Gravitational Lensing Experiment (OGLE), the MAssive Compact Halo Object (MACHO) project, \textit{Gaia}'s Data Release 2, and in some cases the All-Sky Automated Survey (ASAS). From an initial sample of 2315 sources, our method provides an unprecedented sample of 1303 LMC  Classical Cepheids with accurate PCRs, the largest for any single galaxy, including the Milky Way. The derived PCRs are largely compatible with theoretically expected values, as computed by our team using the Modules for Experiments in Stellar Astrophysics (MESA) code, as well as with similar previous computations available in the literature. Additionally, five long-period ($P>50\,\rm{d}$) sources display a cyclic behavior in their $O-C$ diagrams, which is clearly incompatible with evolutionary changes. Finally, on the basis of their large positive PCR values, two first-crossing Cepheid candidates are identified. 
\end{abstract}




\section{Introduction} \label{sec:intro}

Classical or Type I Cepheids (hereafter Cepheids) are pulsating stars named after $\delta$ Cepheid, one of the first variable stars of this kind to be discovered (the first one actually being $\eta$ Aquilae; \citeauthor{Catelan2015}, \citeyear{Catelan2015}). Owing to their period-luminosity relationship \citep[also known as the {\em Leavitt Law};][]{Leavitt1912} and intrinsic brightness, Cepheids play a key role in constraining the Hubble constant by working as standard candles \citep[e.g.,][and references therein]{Freedman-2001,Breuval2020,Riess2021}. 

\par As they are intermediate-mass stars, their evolution is relatively fast and their ages are on the order of $10^7$ and $10^8$ yr for the longest-period (also the most massive and most luminous) and shortest-period (the least massive and least luminous) Cepheids, respectively. Their pulsations are excited by the $\kappa$ and $\gamma$ mechanisms, with the zone where helium goes from singly to doubly ionized being the most important for driving pulsations \citep[see, e.g.,][and references therein]{Catelan2015}.

\par Stellar evolution theory predicts three different stages where these stars become unstable to radial pulsations as they cross the instability strip in the Hertzsprung-Russell diagram (HRD): once during hydrogen shell burning and twice during core helium burning \citep[e.g.,][]{Turner2006}.
According to the Stefan-Boltzmann law, 
a change in effective temperature at approximately constant luminosity implies a change in radius. Moreover, as the period-mean density relation \citep[also known as {\em Ritter's relation};][]{Ritter1879} is valid within the instability strip, changes in radius result in changes in period. Such period changes have been previously detected by numerous authors, and they seem to be more stable in Cepheids pulsating in the fundamental mode \citep{Poleski2008}.

\par Period changes are useful for probing stellar evolution. Examples in the literature include, among many others,  \citet{Pietrukowicz2001}, \citet{Turner2006}, and \citet{Karczmarek2011} for Cepheids, and \citet{LeBorgne2007}, \citet{Silva-Aguirre-2008},  and \citet{Jurcsik-2012} for RR Lyrae stars. 
Since theory predicts that the timescales for these evolutionary changes, $P\,\dot{P}^{-1}$, fall in the range $10^4-10^7$~yr \citep[e.g.,][]{Poleski2008}, observations covering several decades are required to compute the associated period change rates (PCRs). A common technique that is used to conduct these measurements involves the so-called {\em observed minus computed} (O-C) diagrams \citep[see][for reviews and additional references]{Zhou1999,Sterken2005}. 
In this sense, the data for Cepheids in the Large Magellanic Cloud (LMC) that has been made available by the Digital Access to a Sky Century @ Harvard project \citep[DASCH;][]{Grindlay2012} are suitable for the computation of PCRs, as they cover approximately a century. Modern CCD photometric surveys extend the baseline provided by DASCH even further, allowing the determination of more accurate PCRs than using DASCH alone.

\par Our goal in this paper is to calculate PCRs for LMC Classical Cepheids pulsating in the fundamental mode, using both DASCH data and more recent datasets from the literature. This also allows us to revisit previously published PCR values calculated for smaller Cepheid samples \citep{Pietrukowicz2001,Karczmarek2011}, and to compare the measured values with those computed on the basis of state-of-the-art theoretical Cepheid models \citep[e.g.,][Espinoza et al. 2021, in preparation]{Turner2006}.

\par This paper is structured as follows: the utilized data sets are described in Section~\ref{sec:Data}, and the procedure to determine the PCRs therefrom is detailed in Section~\ref{sec:Method}. The main results are presented in Section~\ref{sec:Results}, and the main conclusions drawn therefrom are summarized in Section~\ref{sec:Conc}. 

\section{Data}\label{sec:Data}
DASCH data play a crucial role in our computation of PCRs due to the time coverage of the observations included therein. We specifically make use of DASCH DR6 and, in order to further extend the time baseline coverage, data from other optical surveys were included, favoring observations obtained in the bands closest to that of the DASCH photometry \citep[roughly Johnson $B$;][]{Tang2013}. The surveys used to extend the time baseline are the MAssive Compact Halo Object Survey \citep[MACHO;][]{Alcock2000}, the Optical Gravitational Lensing Experiment \citep[OGLE;][]{Udalski2015}, and \textit{Gaia}'s Data Release 2 \citep[DR2;][]{GaiaCollaboration2018,GaiaDR2-phot2018}. 
Additionally, the longest-period, brightest Cepheids were usually saturated in the OGLE and MACHO surveys, therefore the DASCH data was augmented by the OGLE Shallow survey \citep{Ulaczyk2012} and, when possible,
the All-Sky Automated Survey \citep[ASAS-3;][]{Pojmanski2002}, as replacements for the saturated OGLE photometry. A summary of the basic properties of these sources of data is provided in Table~\ref{tab:srvs}.

\begin{table*}
\caption{Data Summary. List of data sources used and their corresponding photometric properties. Dates are taken from the values available in the collected data set and rounded to the nearest integer year. Note that DASCH data include two different calibrations, one being the APASS $B$-band calibration (1A) and the other the GSC2.3.2 calibration (1B).}
\label{tab:srvs}
\begin{tabular}{llcccl}
\hline
{Data} & {Code} & {Band}  & {Date} & {Reference}\\ 
 & & & {yr} &\\
\hline
DASCH         & 1A, 1B     & Photographic  & $1890-1990$ & \cite{Tang2013} \\
MACHO         & 2     & MACHO $b$             & $1993-2000$ & \cite{Alcock2000} \\
OGLE-III      & 3 & Johnson $V$             & $1997-2008$ & \cite{Soszynski2008} \\
ASAS-3        & 4     & Johnson $V$             & $2001-2010$ & \cite{Pojmanski2002} \\
OGLE Shallow  & 5   & Johnson $V$             & $2005-2009$ & \cite{Ulaczyk2012} \\
OGLE-IV       & 6  & Johnson $V$             & $2010-2016$ & \cite{Udalski2015} \\
\textit{Gaia} DR2 & 7 & {\em Gaia} BP & $2015-2016$ & \cite{GaiaCollaboration2018} \\
\hline
\end{tabular}
\end{table*}

\section{Method}\label{sec:Method}

\subsection{Source Selection}
By using the OGLE Collection of Variable Stars \citep[OCVS;][]{Soszynski2015}, 2315 Cepheids pulsating in their fundamental mode were selected and their periods retrieved. Owing to the characteristics of the data in DASCH, the sample was split: While brighter and longer period sources often have clear and well-sampled light curves, fainter and shorter period sources tend to have incomplete phase coverage and show problems (such as blending) more frequently. Therefore, two subsets of data were utilized: 242 sources with OGLE $V<15$ and 2073 fainter sources.

\par As mentioned in Section~\ref{sec:Data}, long-period sources ($P>50\,\rm{d}$) were saturated in OGLE and therefore they would not have measured magnitudes in the OCVS, although their periods are available. As a result, they were missing from our initial sample and had to be added manually. This added 5 sources to our sample.

\par Counterparts in MACHO (and ASAS, when necessary) were found by using the crossmatch tool provided with \textsc{topcat} \citep{Taylor2005} and a $3\arcsec$ search radius. In the DASCH case, the sources' respective OGLE IDs can be searched using DASCH's light curve generator tool. Note that sometimes the latter procedure leads to multiple matches, in which case the closest and most complete matching source (i.e., the one with the higher number of observations) was selected.

\subsection{Available Calibrations of the DASCH Survey}\label{daschsel}

DASCH offers several photometric calibration options, from which only two were selected for this work. Analysis of light curves based on the different calibrations showed that generally the AAVSO Photometric All-Sky Survey \citep[APASS;][]{Henden2012} calibration provides a good match with the unfiltered photographic plate blue, but it is limited to an APASS magnitude range between 9 and 15~mag. For fainter sources, the Guide Star Catalog 2.3.2 \citep[GSC2.3.2;][]{Lasker2008} calibration is superior, in terms of quantity of data recovered from the photographic plates and phase coverage of the light curves.

\par Note that there are some sources near the limit of the APASS magnitude range which had data available under both calibrations, and so we had to decide which one to use for the final solution. The selection was based on the number of available points, coverage of the phase-folded light curve, and light curve scatter, the latter two steps done by visual inspection.

\subsection{Data Preparation}

Our method used for the calculation of the PCR values (see Section~\ref{ocmet}) implicitly assumes that the light curves have the same general properties. As this is not strictly true between the different surveys, we standardize the different data sets according to a few rules. 
\par First, the time system of ${\rm HJD} - 2,\!450,\!000$ (modified Heliocentric Julian Date) used by the OGLE survey was adopted for all sources of data. Therefore, conversions were done when necessary, with the only exception of \textit{Gaia} DR2 data, as its modified Barycentric Julian Date (BJD) system was not converted into HJD and instead only changed to ${\rm BJD} - 2,\!450,\!000$. The reason for this is the relatively small expected difference of a few seconds \citep[e.g.,][]{Seidelmann1992} between HJD and \textit{Gaia} BJD times of observations, which is irrelevant given the much longer pulsation periods of Cepheids.

\par In addition, in order to properly combine different data sets, a common zero-point and amplitude had to be enforced. This was done by subtracting the median magnitude from each data set and by scaling the magnitudes from each survey so they would have a common standard deviation, set to $\sigma = 0.3$. This value is arbitrary, and was chosen purely because it allows us to recover approximately the amplitudes of large-amplitude Cepheids in the $V$-band. However, due to the specifics of our $O-C$ method implementation, the choice of $\sigma$ has no effect on the final derived period-change rates (as verified on a few variables by varying it in the range $0.1-1.0$). The method involving standard deviation was chosen over light curve normalization by amplitude due to the presence of outliers, especially in the case of DASCH light curves.

\subsection{The Adopted O-C Method}\label{ocmet}

The points in the $O-C$ diagram are derived from the timing differences between observed times of a specific point in the light curve, usually the maximum or minimum value, and the corresponding values computed (or expected) assuming a constant period. In this work, following \cite{Her1919}, we use the entire light curve morphology to compute these differences. A light-curve template is constructed through a truncated Fourier series (TFS) as follows:

\begin{eqnarray}\label{eq:fourier}
m^* & \approx & A_0+\sum^n_{i=1}\left[A_i\cos\left(2\pi i t^*\right)+B_i\sin\left(2\pi i t^*\right)\right], \\
t^* & = & \frac{t}{P_{\rm OGLE}},
\end{eqnarray}

\noindent where $m^*$ are the magnitudes of a section of the light curves used for the construction of the template, $t$ are the modified HJD times, $P_{\rm OGLE}$ is the OGLE period, $n$ is the number of terms in the series, and $A_{0}$, $A_{i}$, and $B_{i}$ are the parameters to be computed. This was done by using an ordinary least squares regression. We have conducted tests to determine the order of the TFS most appropriate to represent the Cepheid light curves. Visual inspection revealed that in most cases, $n = 10$ is needed to properly represent the light curve (especially near the often rather steep rising branch). 
\par The templates are constructed by using OGLE-IV data. When no OGLE-IV data are available, OGLE-III is used instead. To compare with the template light curve, data were split into bins of width 350 days, except for the DASCH data which, due to the higher scatter of the photographic light curves, generally require more data points, and hence wider bins, to measure $O-C$ values accurately. Therefore, in the latter case we applied bin widths of 1400 days.

\par The $O-C$ values were determined for all bins containing more than 10 points by minimizing the scatter between the template and the data in each of the bins. In practice, the data within each of the bins was folded with the adopted period, and the residuals minimized with a non-linear least-squares method, where the only variable is a shift applied to the times of the light-curve points. We estimated the $O-C$ values sequentially, starting with the most recent data and proceeding backwards in time, always using the last $O-C$ value estimated as the starting point in the non-linear optimization of the preceding (in time) bin.

\par We have also determined the uncertainty of each of the measured $O-C$ values using a bootstrapping method. We draw 100 new random data sets from the light curves within each of the data bins, for a number of points equal to the total within the respective bin, with replacement. The $O-C$ values of these bootstrapped samples were also determined, and their scatter with respect to the original $O-C$ value determined for each bin was adopted as the estimate of their uncertainties.

\begin{table*}
\caption{Computed Period Change Rates. This table is published in its entirety in the machine-readable format. The first 10 rows are shown here for guidance regarding its form and content. For each source, from left to right, the information given is as follows: OGLE identifier (following the format OGLE-LMC-CEP-\textit{NNNN}), right ascension, declination, period used to construct the $O-C$ diagram (see Section~\ref{ocmet}), period-change rate and its associated uncertainty (adopted as the mean and standard deviation of the marginalized posterior distribution, respectively), available data sets (see Table~\ref{tab:srvs} for the meaning of these codes), number of Fourier terms for building the light curve and data source used to construct the template (with codes again following Table~\ref{tab:srvs}).}
\label{tab:dpdts}
\begin{tabular}{c.....lll}
\hline
{Source} & \multicolumn1c{$\quad\quad {\rm RA}$ (J2000)} & \multicolumn1c{$\quad\quad {\rm Dec}$ (J2000)} & \multicolumn1c{$\qquad\quad P_{\rm OGLE}$} & 
\multicolumn1c{$\dot{P}$ } & 
\multicolumn1c{$\Delta\dot{P}$} & {Data} & {Fourier Terms} & {Template}\\ 
 & \multicolumn1c{$\quad\quad$(deg)} & \multicolumn1c{$\quad\quad$(deg)} & \multicolumn1c{$\qquad\quad$(d)} & \multicolumn1c{(d Myr\textsuperscript{-1})} & 
\multicolumn1c{(d Myr\textsuperscript{-1})} &  &  & \\
\hline
0005 & 68.881335 & -69.734944 & 5.6119491 & -3.2 & 1.3 & 1B,\,3,\,6,\,7 & 10 & 6 \\
0017 & 69.736785 & -69.688417 & 3.6772904 & -1.6 & 0.4 & 1B,\,3,\,6,\,7 & 6 & 3,\,7 \\
0018 & 69.745755 & -68.956889 & 4.0478369 & 0.0 & 0.7 & 1B,\,3,\,6,\,7 & 6 & 6 \\
0021 & 69.879090 & -69.261833 & 5.4579579 & 5.3 & 1.5 & 1A,\,3,\,6,\,7 & 10 & 6 \\
0025 & 70.226535 & -69.115083 & 3.7334902 & 0.3 & 0.7 & 1B,\,3,\,6,\,7 & 10 & 6 \\
0027 & 70.331160 & -67.335583 & 3.5229124 & -1.9 & 0.3 & 1B,\,3,\,6,\,7 & 10 & 6 \\
0033 & 70.491375 & -69.433556 & 7.1807603 & 7.4 & 1.3 & 1A,\,3,\,6,\,7 & 10 & 6 \\
0034 & 70.507965 & -70.646944 & 11.2552749 & 68.4 & 9.1 & 1A,\,3,\,6,\,7 & 8 & 6 \\
0035 & 70.526295 & -67.969889 & 6.9435423 & 3.4 & 1.8 & 1A,\,3,\,6,\,7 & 8 & 6 \\
0039 & 70.658370 & -68.516472 & 3.1477882 & -0.1 & 0.6 & 1B,\,3,\,6 & 6 & 6,\,7 \\
\hline
\end{tabular}
\end{table*}

\par Having computed the $O-C$ values, it is straightforward to build the $O-C$ diagrams. In these, the $x$-axis corresponds to a representative time for each bin (mean of the bin edges), while the $y$-axis is the found $O-C$ value. Assuming that the change in period is linear, we can expect a parabola in the diagram. The quadratic coefficient of this parabola is then related to the PCR \citep[e.g.,][]{Sterken2005} according to the following equation:

\begin{eqnarray}\label{eq:dpdt}
O-C = \frac{1}{2}\frac{dP}{dt}\overline{P}E^2,
\end{eqnarray}

\noindent with $E$ the epoch and $\overline{P}$ the average period. For the latter, we have used the periods exactly as given in the OCVS, as the absolute period change values are small in the case of Cepheids. Also, while the exact period value does change the appearance of the $O-C$ diagrams, it has a negligible effect on the measured period-change rates~-- i.e., on the order of seconds per year, as shown by observations and theoretical models alike \citep[][]{Turner2006}. Therefore, in our computations we adopt $\overline{P} \equiv P_{\rm OGLE}$.

\par To find the quadratic trend in the data and relate it to equation \ref{eq:dpdt}, a least-squares method which includes the uncertainties from each $O-C$ value as weights was used. This first result was taken as the starting point for the Markov chain Monte Carlo (MCMC) modeling. We used the \textsc{emcee} package \citep{ForemanMackey2013}, which implements an affine invariant ensemble method for the calculation of the posterior distribution of parameters. The logarithmic likelihood was adopted as follows:

\begin{eqnarray}
\eta & = & aE^2 + bE + c, \\
\mathcal{L} & = & -\frac{1}{2} \sum_j\left[\left(\frac{y-\eta}{y_e}\right)^2 + \log\left(y_e^2\right)\right],
\end{eqnarray}

\noindent where $\eta$ is the equation that describes the quadratic model, $(a,b,c)$ are the coefficients of the quadratic polynomial fit, $\mathcal{L}$ is the logarithmic likelihood, $j$ represents each of the bins of the $O-C$ diagram, $y$ are the $O-C$ values, and $y_e$ the $O-C$ uncertainties (standard deviations from the bootstrap). In this work, the \textsc{EMCEE} code was run using 32 walkers for 10000 steps, of which 5000 is the burn-in period, and a thinning factor of 15. For each of the parameters, the means and standard deviations of their marginalized posterior distributions are adopted as their final estimated values and uncertainties, respectively.

\par At this point, some difficulties were found and specific corrections were applied, as follows: 

\renewcommand{\labelenumi}{\roman{enumi}.}
\begin{enumerate}
    \item Due to the periodicity of the TFS and lack of data in some epochs, a shift in some $O-C$ values equal to $kP$ (with $k \in \mathbb{Z}$) was found. To fix this issue, a visual inspection and the required displacement equal to $-kP$ were executed.
    
    \item Some variables have little or no published $V$-band photometry in OGLE. In these cases, the light-curve templates were constructed using either \textit{Gaia}, a combination of OGLE (when available) and \textit{Gaia}, or MACHO light curves.
    
    \item Some variables lack sharp features in their light curves. This sometimes results in unrealistic template shapes, when prepared using the default 10 Fourier terms. For these, the number of terms were reduced upon visual inspection. The changes are specified on table \ref{tab:dpdts}.
    
    \item Some \textit{Gaia} light curves have poor phase coverage, resulting in scaling and zero-point problems. These were manually scaled after visual inspection.
    
    \item Some sources were blended or would not have a distinguishable light curve in the DASCH data. Most of these required data selection using the DASCH quality flags.
    
    \item For the second subset (dim sources), blending, inadequate phase coverage and noise made the data preparation fail frequently. Therefore, an additional requirement of having more than 200 data points in DASCH as well as a period longer than 3 days was imposed on this subset after visual inspection of the preliminary results. Additionally, the magnitude scaling was changed from an automatic process to a semi-automatic one, allowing us to modify slightly the initial amplitude and zero-points when necessary, for a better agreement between data from different surveys.
\end{enumerate}

The final results were obtained after implementing these changes to the fitting process, when necessary.

\section{Results}\label{sec:Results}
We have successfully measured period-change rates for 1303 variables out of our initial sample of 2315. These values, alongside other supporting information, are listed in Table~\ref{tab:dpdts}.

\par
\begin{figure*}
    \includegraphics[width = 2\columnwidth]{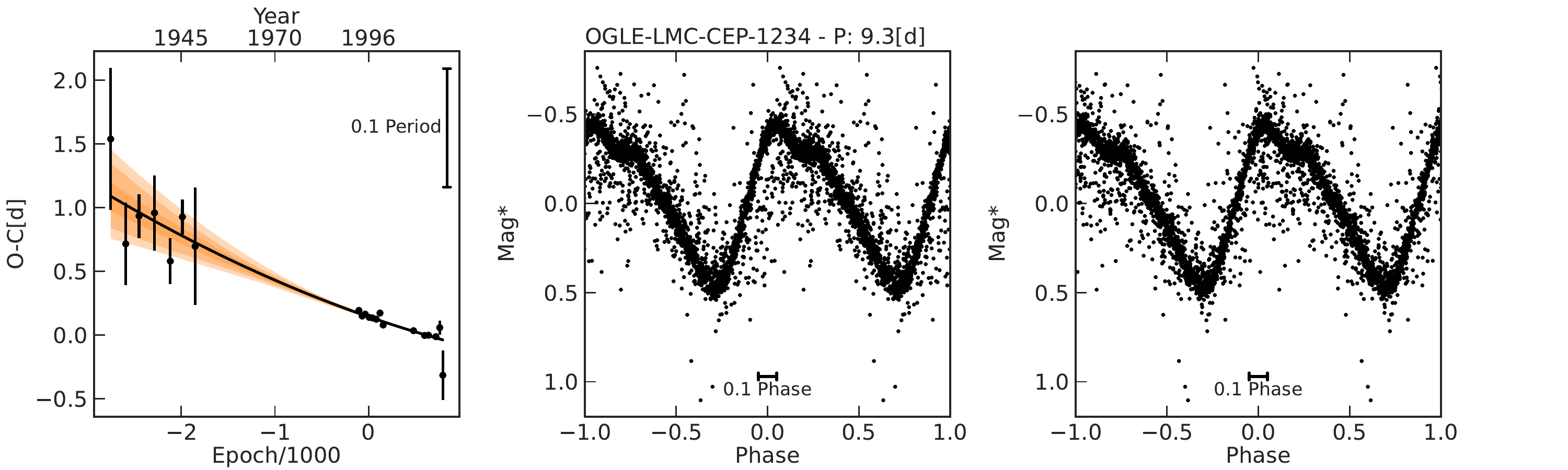}
    \includegraphics[width = 2\columnwidth]{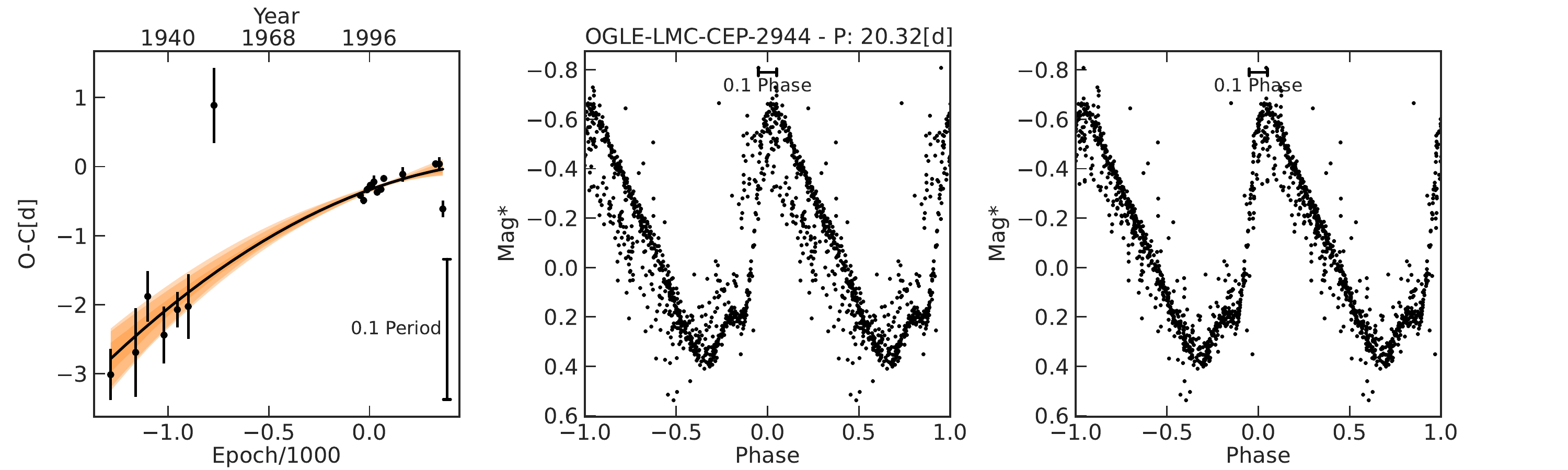}
    \includegraphics[width = 2\columnwidth]{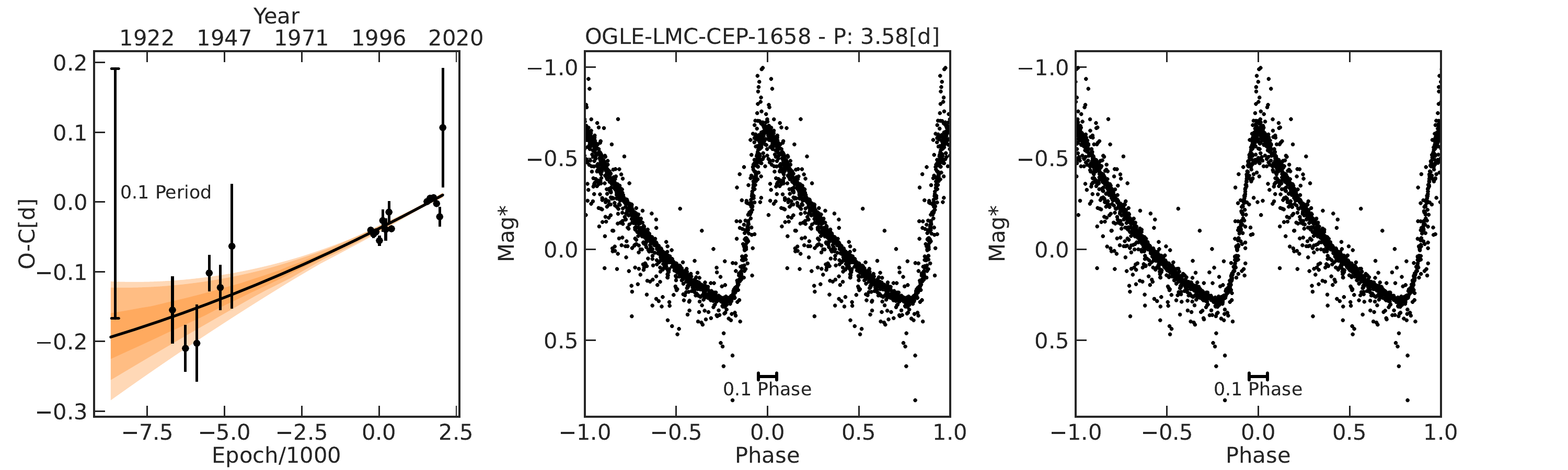}
    \caption{Examples of $O-C$ diagrams from our sample of LMC Cepheids with computed period changes. The \textit{top, middle, and bottom} panels illustrate cases with positive, negative, and marginal period-change rates, respectively. The \textit{left} panels show our measured $O-C$ values, the ensemble of MCMC solutions (shaded areas show the ranges where posterior solutions corresponding to one, two, and three standard deviations are found, in order of decreasing transparency), and the adopted final period-change solution (continuous black parabolas). The \textit{middle} and \textit{right} panels exhibit the folded light curves, uncorrected ({\em middle}) and corrected ({\em right}) for the computed period changes. Note the improvement in the cases of positive (\textit{top}) and negative (\textit{middle}) period changes, when compared to the marginal improvement in the small period-change case (\textit{bottom}). Phase on the x-axis stands for an arbitrary pulsation phase and Mag$^*$ on the y-axis stands for the magnitudes modified as stated in Section~\ref{sec:Method}. Scale bars are plotted in each panel for reference.}
    \label{fig:0}
\end{figure*}

 Figure~\ref{fig:0} presents examples of $O-C$ diagrams and folded light curves (both before and after correcting for our measured PCR) for stars with positive, negative, and very small (consistent with zero) period changes.

\subsection{First Crossing Candidates}\label{sec:firstc}

First-crossing Cepheids are very rare objects, as they are evolving very rapidly towards lower effective temperatures after crossing the Hertzsprung gap on the HRD. We detected two such candidates, namely OGLE-LMC-CEP-2840, with a period of $3.8\,\rm{d}$ and a PCR of $30.9\,\rm{d\,Myr}^{-1}$, as well as OGLE-LMC-CEP-2132, which has a period of $4.68\,\rm{d}$ and a PCR of $179.8\,\rm{d\,Myr}^{-1}$. In spite of their likely different metallicities, these results are comparable to what was found by \cite{Kov2019} for the Galactic Cepheid V1033~Cyg, which has a reported period of $4.9\,\rm{d}$ and a PCR of $210.5\pm0.9\,\rm{d\,Myr}^{-1}$. 

\par
\begin{figure*}

\includegraphics[width = 2\columnwidth]{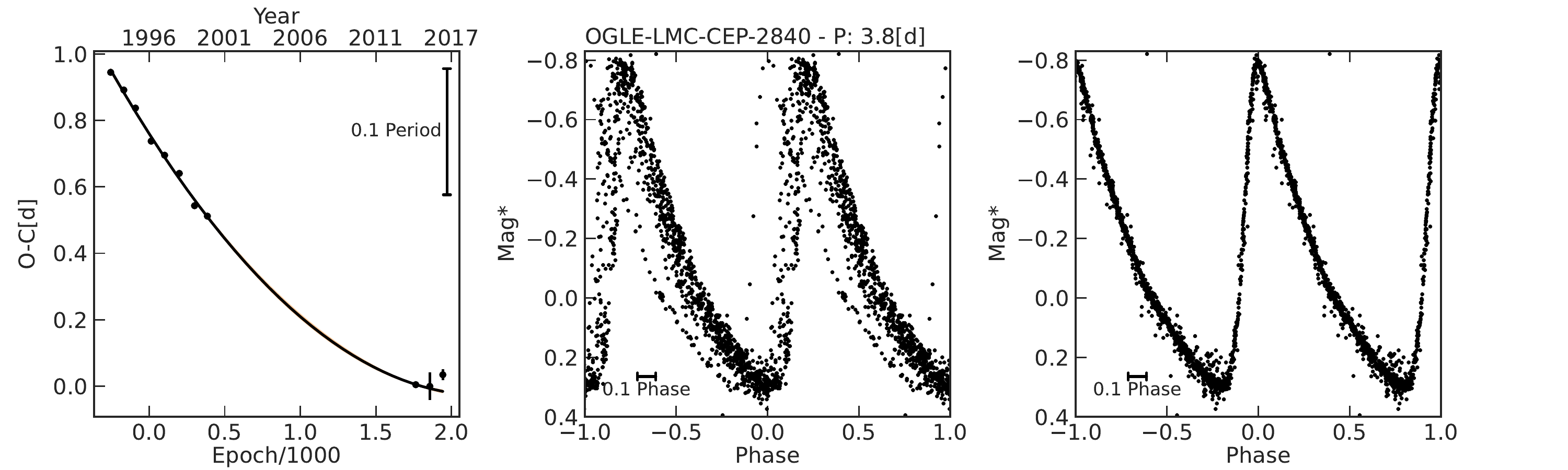}

\includegraphics[width = 2\columnwidth]{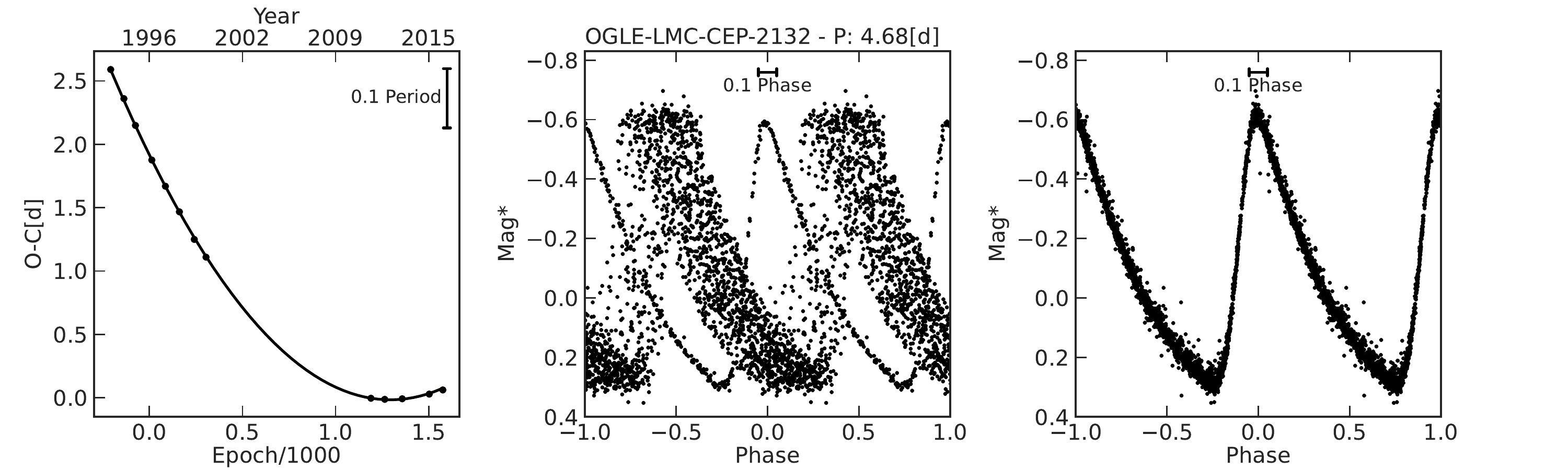}

\includegraphics[width = 1.2\columnwidth]{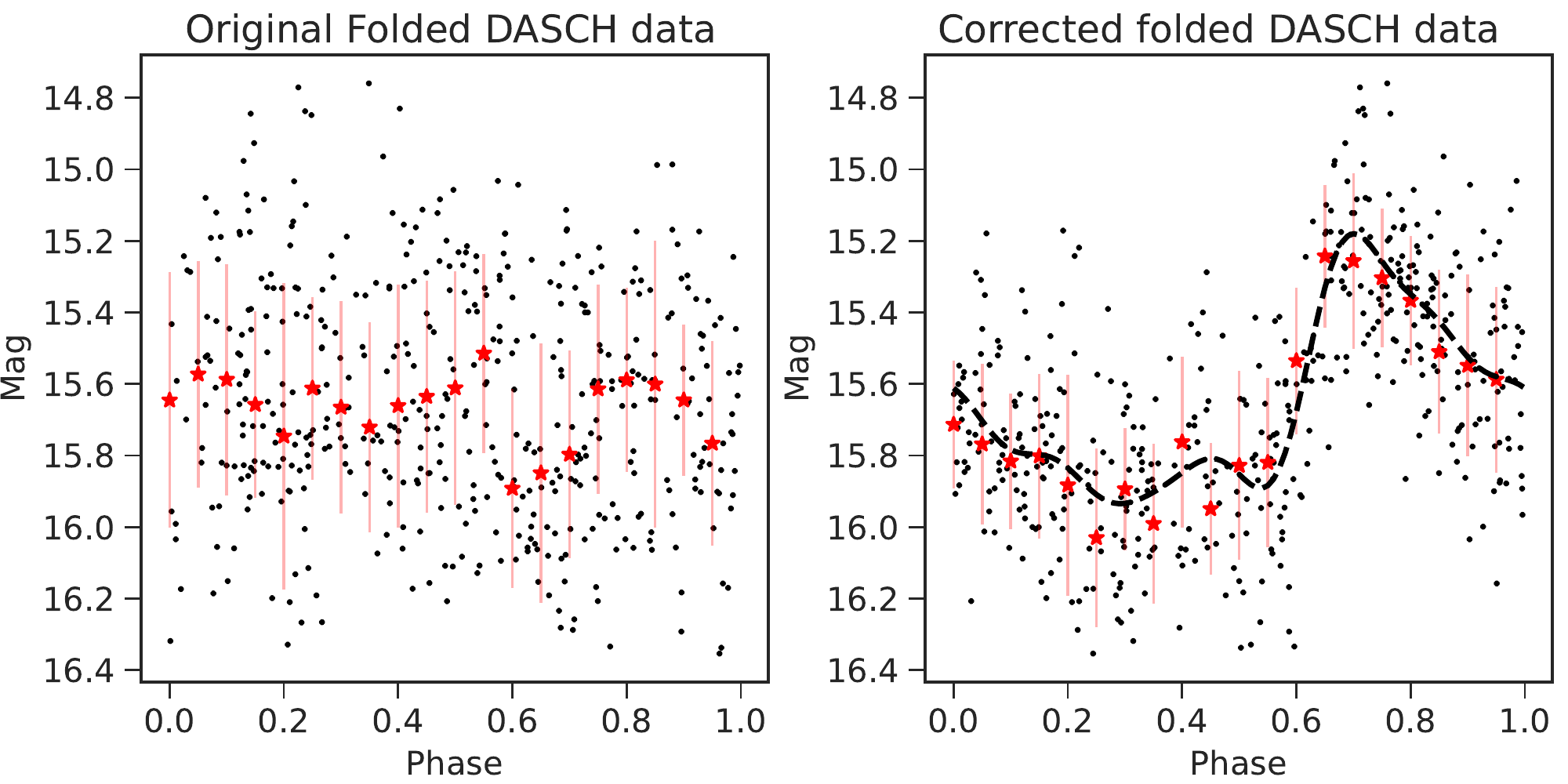}

\caption{First-crossing Cepheid candidates. The \textit{top} and \textit{middle} panels show the two candidates in our sample, with each panel having the same meaning as the corresponding ones in Figure~\ref{fig:0}. Here too the MCMC ensemble results are overplotted, but they are virtually imperceptible because of the very small inferred errors. For both sources, DASCH data were available, but were not used for the determination of the PCR due to the cycle number ambiguity (see Section~\ref{ocmet}). The \textit{bottom left} panel shows the original DASCH data for OGLE-LMC-CEP-2132 folded with $P_{\rm OGLE}$, while the \textit{bottom right panel} is similar, but after correcting for the period changes inferred from the corresponding $O-C$ diagram. The data were binned in phase intervals of 0.05 phase, from which the mean and standard deviation were computed and plotted as red markers and error bars, respectively. The dashed line in the latter plot, built with a sixth-order Fourier series computed using the robust Huber loss function, shows the approximate light-curve shape.}
\label{fig:first}
\end{figure*}

 The $O-C$ diagrams and folded light curves for the new first-crossing candidates are shown in Figure~\ref{fig:first}. These sources were found upon visual inspection, as the algorithm could not account properly for the changes that occurred between the DASCH observations and more recent ones. For this reason, the $O-C$ diagrams were prepared, and the PCRs determined, without the DASCH data. 

\par Nevertheless, in the case of OGLE-LMC-CEP-2132, the fit to the recent data allowed us to correct the DASCH observations for the effect of period changes and fold those, hence validating our approach. In the case of OGLE-LMC-CEP-2840, the scatter of the DASCH data prevents us from doing the same, unfortunately. As a further sanity check, we have also constructed the $O-C$ diagrams using the light curves obtained with the redder bands available for each data set (OGLE's $I$, {\em Gaia}'s RP, and MACHO's $r$). The inferred PCR values, $175.2\,\rm{d\,Myr}^{-1}$ and $48 \,\rm{d\,Myr}^{-1}$ for OGLE-LMC-CEP-2132 and OGLE-LMC-CEP-2840, respectively, are similar to those previously obtained for these stars, which provides further support for our analysis technique.

\par These variables are interesting targets for spectroscopic studies, as first-crossing Cepheids are expected to have surface abundances unaffected by the first dredge-up on the red giant branch \citep{Kov2019}, as they have not yet reached this phase in their evolution. In particular, Cepheids with lithium overabundances are thought to be in their first crossing \citep[e.g.,][and references therein]{Luck2001,Kov2019,Catanzaro2020,Ripepi2021}, and thus a determination of the Li abundances in our two candidates would be of particular interest.

\subsection{Comparison with Theoretical Models and Galactic Cepheids from Turner et al. (2006)}\label{sec:models}

\begin{figure*}
\includegraphics[width = 2\columnwidth]{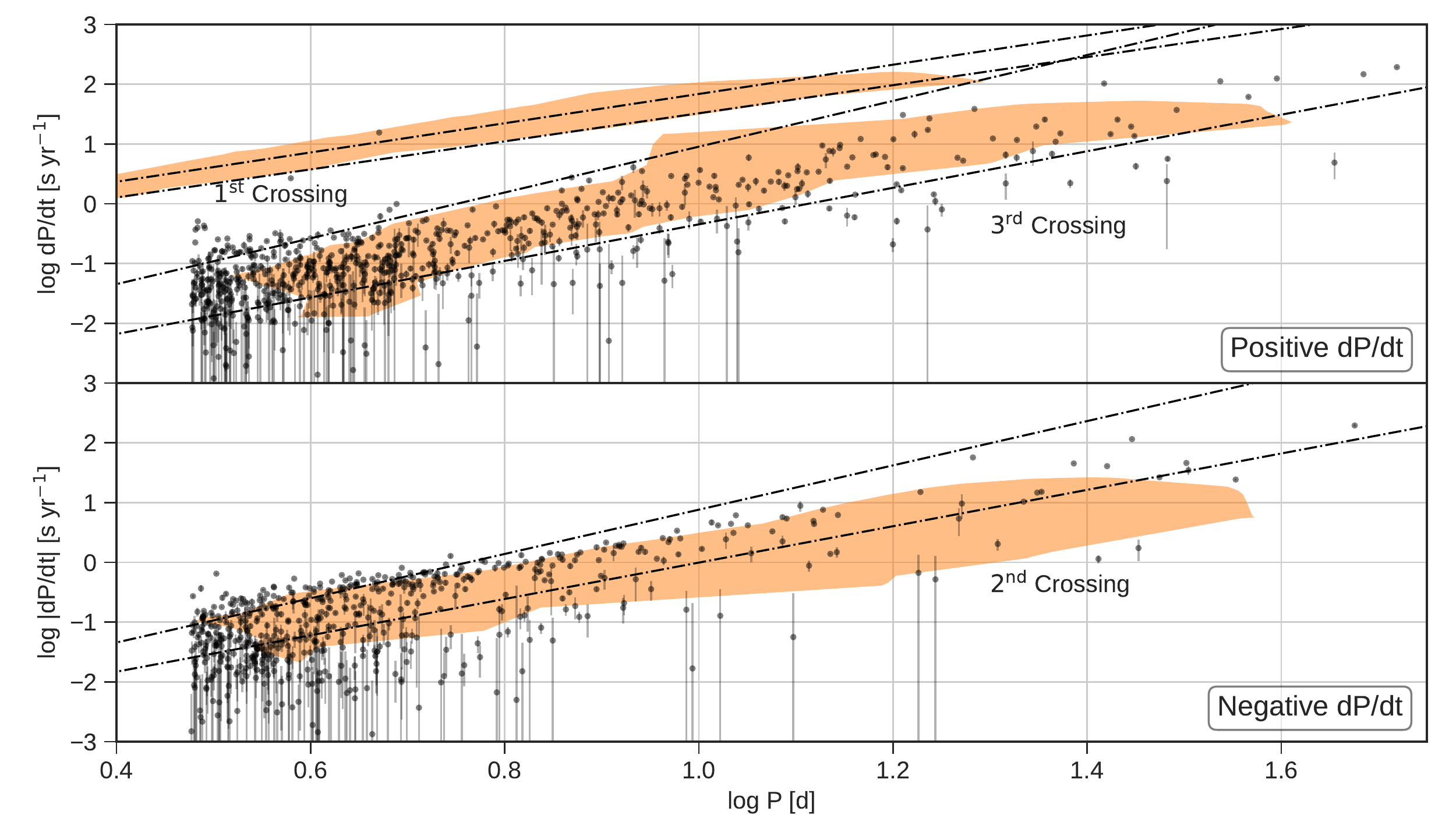}
\caption{The dependence of measured positive (\textit{top} panel) and negative (\textit{bottom} panel) PCRs on the pulsation period. The dash-dotted lines are the regions of expected PCRs for the first, second, and third crossings of the instability strip for the set of models adopted in \protect\cite{Turner2006} (their Fig.~2). The shaded areas are new computations (Espinoza-Arancibia et al. 2021, in preparation) using the single-star evolution and pulsation modules of MESA \citep{Paxton2019} (see Sect.~\ref{sec:models}).}
\label{fig:1}
\end{figure*}

\cite{Turner2006} presented a detailed comparison between PCRs measured for Galactic Cepheids and the values expected from models. We present the dependence of the measured PCRs on the pulsation periods in Figure~\ref{fig:1}, following the example of \cite[][their Fig.~2]{Turner2006}. Although not all values fall within the expected areas for crossings according to \cite{Turner2006}, we found 571 and 700 PCRs consistent with Cepheids on their second and third instability strip crossing according to their negative or positive PCR values, respectively. This implies an approximate ratio of $4:5$ between the number of sources with $P > 3$~d in the second and third crossings. The remaining Cepheids include 30 sources that are consistent with no period changes plus two first-crossing candidates with large PCRs (Sect.~\ref{sec:firstc}). 

\par In order to carry out a more systematic comparison with model predictions, we computed evolutionary models both with and without rotation for Classical Cepheids, using the Modules for Experiments in Stellar Astrophysics (MESA) code \citep{Paxton-2011}. The pulsation analysis of the thus computed models was carried out using the Radial Stellar Pulsation (RSP) module, which is also integrated into MESA \citep{Paxton2019}. A full description of these calculations is beyond the scope of this paper, and will be provided in Espinoza-Arancibia et al. (2021, in preparation).

In Figure~\ref{fig:1} we show the results of our calculations in the $\log (dP/dt)$ vs. $\log P$ plane, for a grid of Cepheid models of metallicity $Z = [0.005,\,0.007,\,0.009]$ and initial masses between $4\, {\rm M}_\odot$ and $7\,{\rm M}_\odot$, uniformly spaced in mass with a resolution of $0.5\,{\rm M}_\odot$. Initial rotation was assumed to be solid body-like and values were taken in steps of $0.1$ within the range $\Omega / \Omega_{\rm crit} \in [0.0,\,0.9]$, where $\Omega$ and $\Omega_{\rm crit}$ are the angular velocity and its critical value, respectively. The loci occupied by these models is schematically indicated as shaded areas, whereas the lines represent the results from \citet{Turner2006}. This figure shows that the predictions of these two sets of models are largely consistent (see Espinoza-Arancibia et al. 2021 for more details).  

\par Note that there is also a good general agreement between the measured PCRs and their expected values from the models. This is observed over the full period range covered by  the Cepheids in our sample, although our adopted low-mass limit does not allow us to draw strong conclusions regarding the behavior of the shortest-period Cepheids, as those are also the ones with the lowest masses. Similarly to \cite{Turner2006}, who only detected one first-crossing candidate in their sample, we confirm that such stars are indeed rare, with only two candidates being present among 1303 stars (see Sect.~\ref{sec:firstc}).

\subsection{Comparison with OGLE LMC Cepheids from Poleski (2008)}
\cite{Poleski2008} also carried out PCR measurements for LMC Cepheids, using data from the OGLE and MACHO projects, and compared his empirical results to the same \cite{Turner2006} models that were discussed in the previous subsection (see his Fig.~3). While many of his Cepheids do indeed fall within the expected region of the $\log (dP/dt)$ vs. $\log P$ plane according to both the \cite{Turner2006} and our models, the majority are placed outside it. In fact, there is even a large number of Cepheids in \cite{Poleski2008} study whose position in this diagram would suggest that they are actually in their first crossing of the instability strip. Those results are in contrast with both our findings and those of \cite{Turner2006}. Note that there are no tables with PCRs computed in \cite{Poleski2008}, and therefore our analysis is qualitative.

\par
\begin{figure*}
\includegraphics[width = 2.1\columnwidth]{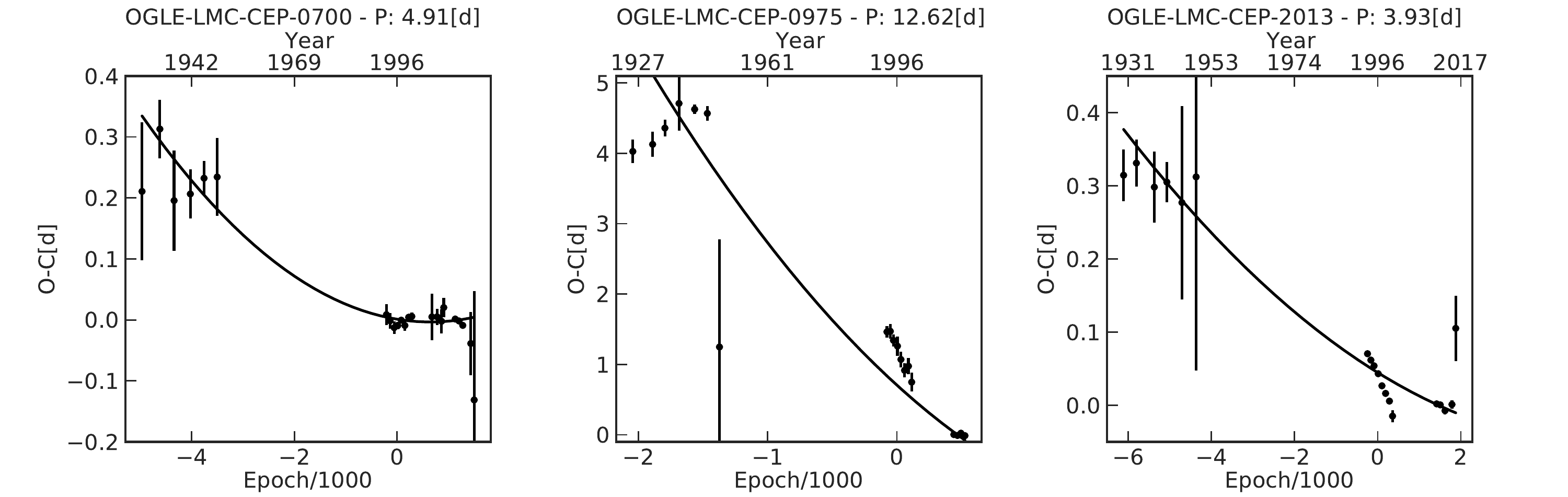}
\caption{Examples of $O-C$ diagrams for Cepheids of short (left and right panels) and intermediate (middle panel) pulsation periods. The presence of non-linear period-change behavior is readily apparent. To highlight this behavior, no MCMC ensembles are shown.}
\label{fig:fluct}
\end{figure*}

 It is important to note that, while \cite{Poleski2008} also used Fourier methods to compute period changes, a key difference lies in the amount of data, and the corresponding time coverage, upon which his measurements are based: \cite{Poleski2008} was limited to MACHO, OGLE-II and OGLE-III data, and the relatively short implied time-span, of roughly 15~yr (compared with a typical 100~yr, in the case of our study), makes it difficult to properly measure the small PCRs of classical Cepheids. In addition, \citet{Poleski2008} carried out individual $O-C$ measurements over timescales of days, as opposed to the timescales of years used in our analysis. This implies, as indeed pointed out in this study, that his results are much more sensitive to random period fluctuations. In particular, his short-period sources ($\log P\lesssim 0.8$) seem to be dominated by these, tending to fall outside the areas expected for the different crossings as can be seen in his Figure~3. While we do not find the exact same behavior (as can be seen in our Figure~\ref{fig:1}, which is equivalent to his Figure~3), we do find an increase in the uncertainties in our PCR measurements for Cepheids in the short-period regime. Similarly, visual inspection of the second subset of sources (i.e., those with $V>15\,\rm{mag}$) revealed that many of the sources in the short-period end of our sample ($\log P\lesssim0.7$) showed an erratic behavior that cannot be adequately described by a parabola, contrary to theoretical expectations. This is related to short-period Cepheids being intrinsically fainter and having bigger uncertainties, as can be seen from the typical value of magnitude uncertainty (acquired from the estimated error of the locally corrected magnitude measurement in the DASCH data), which is $0.15$ and increases to $0.6\,\rm{mag}$ for typical Cepheids of magnitude $14$ and fainter than $16$ in the $V$-band, respectively (or, equivalently, $\log P \approx 1.5\,\text{and } 0.5$). While this behavior is more common on these short-period Cepheids, there are some Cepheids with longer periods that show a similar behavior. As an example, the individual $O-C$ values obtained for the Cepheid OGLE-LMC-CEP-0975, with a period of $12.6\,{\rm d}$, have for the most part very small uncertainties, as can be seen in the middle panel of Figure~\ref{fig:fluct}. However, it is also possible to notice in the same panel that there is no clear quadratic trend in the $O-C$ diagram. The other two panels in Figure~\ref{fig:fluct} show additional examples, but in the short-period regime.

\par As mentioned earlier, it is possible that these fluctuations are related to the timescales used. In our analysis, they become more noticeable when the period changes themselves become smaller, which is reflected in the $O-C$ diagrams as $O-C$ values smaller than a day. This could mean that the PCR fluctuations become relevant when they are of similar order as the period changes, which would explain their larger incidence as one approaches the short-period regime. On the other hand, this explanation would not be adequate in the case of long-period Cepheids with erratic behavior, such as OGLE-LMC-CEP-0975. For further discussion about the possible origins of these fluctuations in Cepheids, the reader is referred to \cite{Szabados1983} and \cite{Neilson2014}.

\subsection{Previously Computed Values for PCRs in the LMC}\label{sec:results_B}

\begin{figure*}
\includegraphics[width = 2\columnwidth]{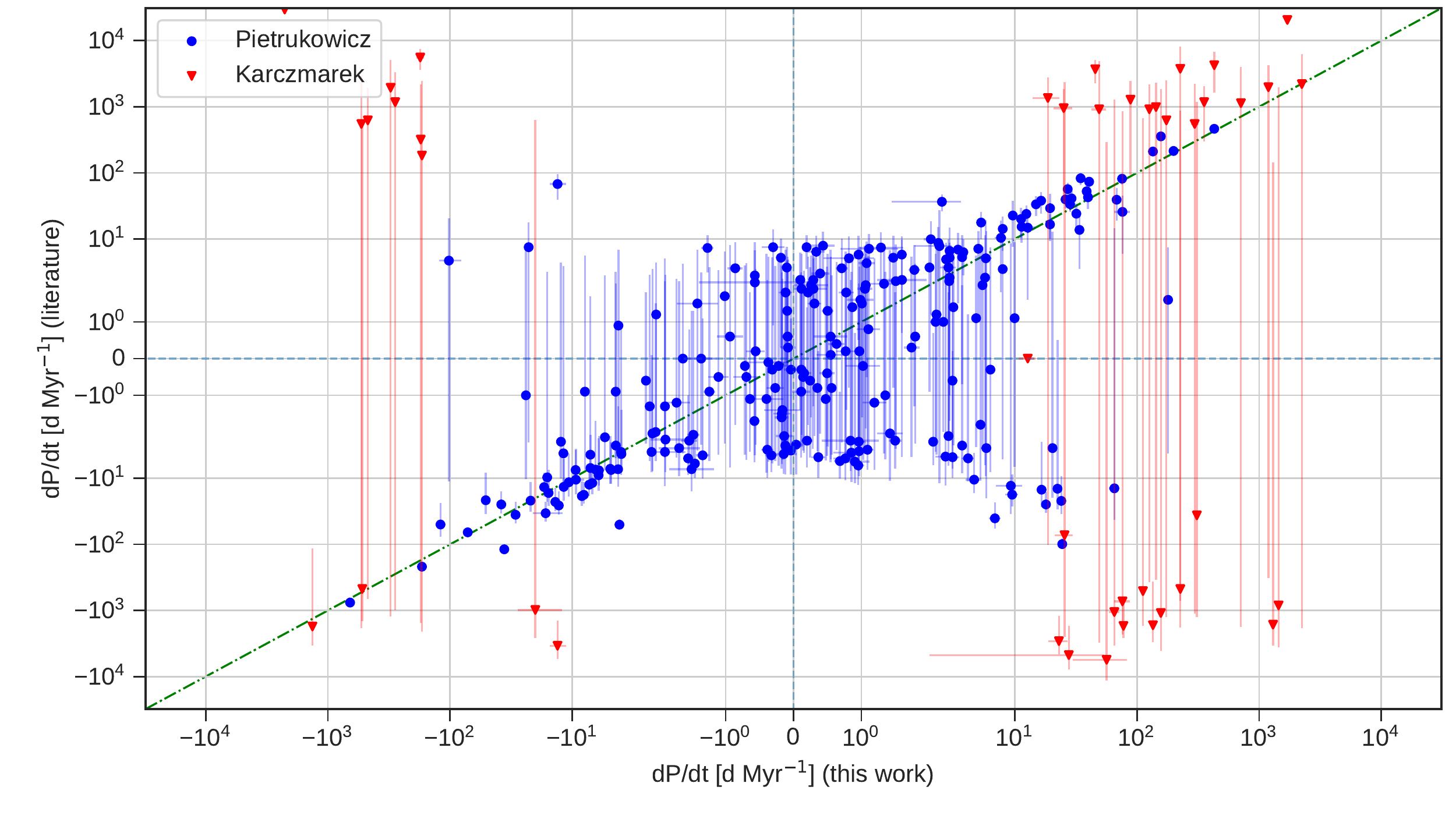}
\caption{Comparison between our computed PCRs for LMC Cepheids and previous results from the literature  \citep{Pietrukowicz2001, Karczmarek2011} for sources in common. The values in the $x$-axis are those from our work, while those in the $y$-axis correspond to either \citet[][blue circles]{Pietrukowicz2001} or \citet[][inverted red triangles]{Karczmarek2011}. The green dash-dotted line shows perfect agreement between previous sets of values and ours ($x = y$), and is plotted for reference only.}\label{fig:3}
\end{figure*}

Both \cite{Pietrukowicz2001} and \cite{Karczmarek2011} computed PCRs for Cepheids in the LMC as well. While \cite{Karczmarek2011} only used ASAS data, \cite{Pietrukowicz2001} employed a data set with a time baseline similar to the one employed in this work, by using the Harvard \citep[as collected by][]{PayneGaposchkin1971}, ASAS, and OGLE data. The results for sources in common between each of those two studies and ours are shown in Figure~\ref{fig:3}. 

\par Close examination of this plot reveals that there is a larger fraction of sources with similar PCRs as measured by us in  \citet{Pietrukowicz2001} than in \citet{Karczmarek2011}. Indeed, the latter's measurements tend to differ from ours by an amount that can reach several orders of magnitude, even in the cases where the sign of their measured period variations is the same as found by us. 

A possible explanation for the differences between our results and those by \citet{Karczmarek2011} could be related to the more limited dataset used by these authors, both in comparison with our work and \citet{Pietrukowicz2001}, and also their focus on variations that occur on shorter timescales. This could lead to their results being dominated by high-frequency noise, as in  \citet[][see the previous subsection]{Poleski2008}. This idea is indeed supported by Figure~8 in \cite{Karczmarek2011}, where a comparison with \cite{Poleski2008} is shown. 

To close, we note that the differences between our results and those by \cite{Pietrukowicz2001} tend to increase with decreasing (absolute) $dP/dt$ values. We have verified that this behavior is primarily caused by the shorter-period Cepheids (our second data subset). This is expected, as data availability and quality in DASCH decreases with decreasing magnitude (period).

\subsection{O-C Diagram Results for Long-Period Cepheids}\label{sec:results_C}

Examples of $O-C$ results for bright Cepheids ($P>50\,\rm{d}$) are shown in Figure~\ref{fig:bright1}. Some properties of these sources are provided in Table~\ref{tab:ldpdts}, where no PCRs are reported, since the computed quadratic trends are not representative of their complex $O-C$ diagrams.

\par Analysis of these plots suggests that, in addition to a possible quadratic component, which could plausibly be related to the long-term, secular evolution of the star, an additional, cyclic component may also be present, which may perhaps be described using a quasi-sinusoidal term \citep[see][his Sect.~4.2]{Sterken2005}. A similar phenomenon is apparent in the $O-C$ diagrams of the two Galactic Cepheids with the longest known periods, namely V1496~Aql, with $P \approx 65$~d, and S~Vul, with $P \approx 68$~d \citep{Berdnikov2004}. The actual mechanism that is responsible for these variations is, however, unclear. A possible explanation might include the presence of one or more companions, which would impact the $O-C$ diagram of the Cepheid by means of the light-travel time effect, as indeed seen in RR Lyrae binaries \citep[e.g.,][]{Hajdu2015, Hajdu-2021}. However, both the amplitude of the variations and the fact that the latter are seen in all our long-period sources, as well as in \mbox{V1496 Aql} and S Vul, make binarity an unlikely general explanation of the phenomenon.

\begin{figure*}

\includegraphics[width = 2\columnwidth]{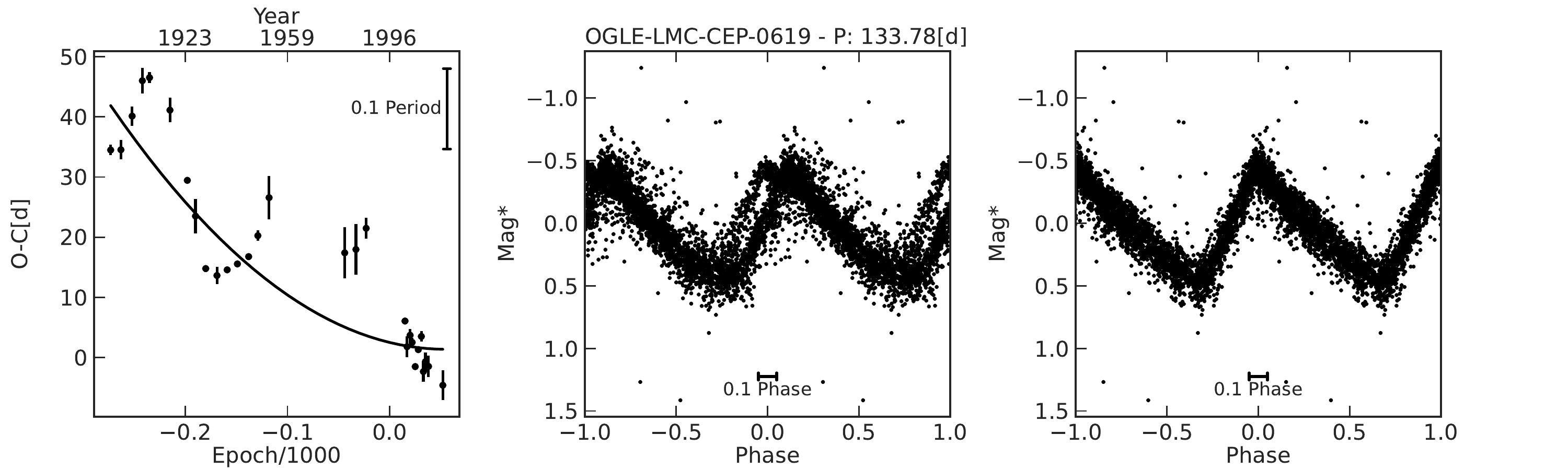}

\includegraphics[width = 2\columnwidth]{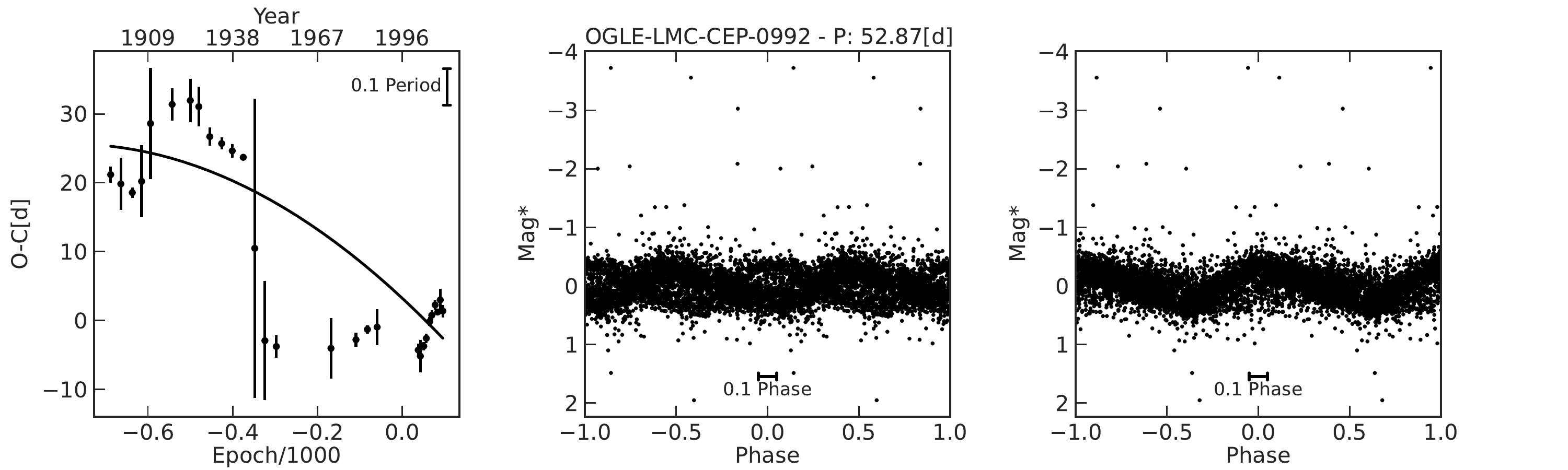}

\includegraphics[width = 2\columnwidth]{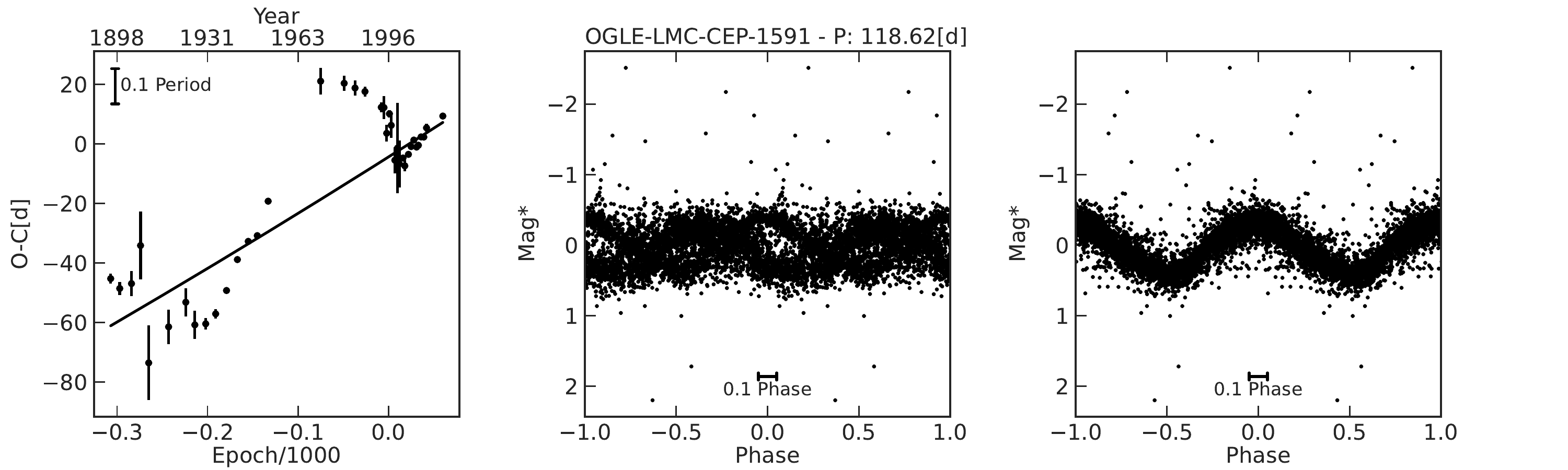}

\caption{For five long-period Cepheids with non-linear period changes, we show their respective 
$O-C$ diagrams (left panels), the original light curves (middle panels), and the light curves obtained after accounting for the inferred period changes, based on the individual $O-C$ values for each epoch (right). Unlike Figure~\ref{fig:0}, the MCMC ensemble results are not shown, as the $O-C$ values do not conform to the parabolic trend that would be expected in the case of linear period changes. The computed parabola is still shown, in order to highlight the complex behavior that is observed.}
\label{fig:bright1}
\end{figure*}

\begin{figure*}
\includegraphics[width = 2\columnwidth]{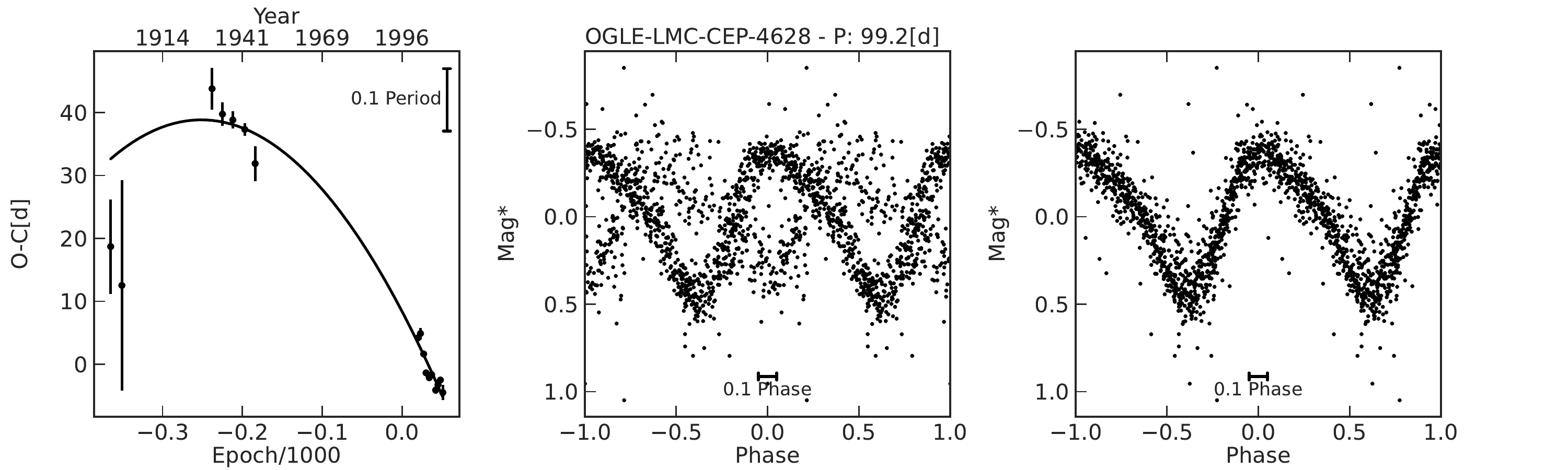}

\includegraphics[width = 2\columnwidth]{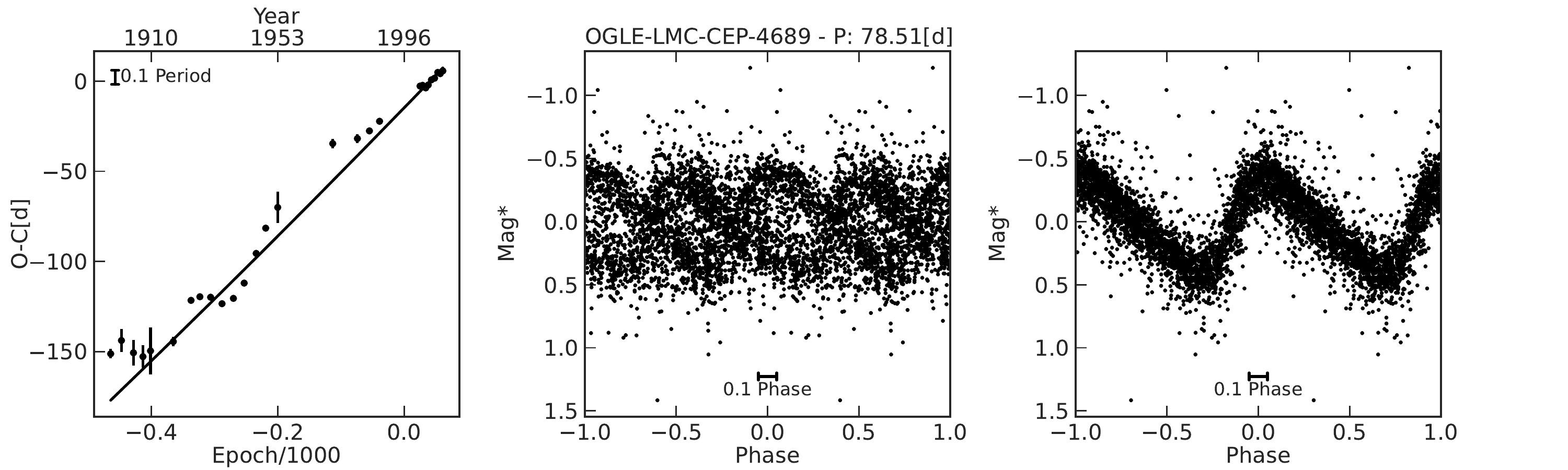}
\contcaption{}
\end{figure*}

\section{Summary and Conclusions}\label{sec:Conc}

\begin{table}
\caption{Data for Long-Period Cepheids. For each long-period Cepheid, the information given are: the OGLE identifier (following the format OGLE-LMC-CEP-\textit{NNNN}), the period used to construct the $O-C$ diagram (see Section~\ref{ocmet}), the available data sets (see Table~\ref{tab:srvs} for the meaning of these codes), number of Fourier terms for the light-curve template and on which survey it was built on.}
\label{tab:ldpdts}
\begin{center}
\begin{tabular}{c.lcc}
\hline
{Source} & \multicolumn1c{$P_{\rm OGLE}$} &{Data} & {Fourier} & {Template} \\ 
{} & \multicolumn1c{(d)} & {} & {} & {}\\
\hline
0619 & 133.78 & 1A,\,4,\,5,\,7 &10& 5 \\
0992 & 52.87 & 1A,\,2,\,4,\,5,\,7 &10& 5 \\
1591 & 118.62 & 1A,\,2,\,4,\,5,\,7 &10& 5 \\
4628 & 99.2 & 1A,\,4,\,7 &6& 4 \\
4689 & 78.51 & 1A,\,4,\,7 &6& 4 \\
\hline
\end{tabular}
\end{center}
\end{table}

Using data from the literature, spanning over 100 years in some cases, we have employed the $O-C$ diagram technique to derive PCRs for an unprecedented sample of 1303 Classical Cepheids in the LMC covering a broad range of periods. Our results were compared with previous results from the literature and with a new set of evolutionary and pulsation models of Classical Cepheids for an LMC-like metallicity and with rotation, computed with MESA. The latter will be presented in more detail in Espinoza-Arancibia et al. (2021, in preparation). Our main conclusions can be summarized as follows.

\par Our model predictions are substantially in agreement with those previously presented by \cite{Turner2006}, over a similar mass range. Both model sets also seem to correctly describe the overall PCR behavior in Classical Cepheids, as brought about by secular stellar evolution. This allowed us to identify two new  first-crossing candidates, for which we suggest spectroscopic follow-up, in order to confirm the presence of typical spectral signatures seen in first-crossing Cepheids, particularly Li enhancement. The majority of the remaining sources have PCR values that are consistent with these Cepheids being in the second or third crossing of the instability strip, as expected.  

\par Comparison with previous PCR measurements show reasonable agreement with \cite{Pietrukowicz2001}, but significant differences with respect to \cite{Karczmarek2011} are present. Neither use an $O-C$ approach and the more limited dataset used by the latter authors (a baseline of only 10~yr), is likely the main reason of the noted differences, given that both a much longer time span was considered in both \cite{Pietrukowicz2001} and our study. As in \citet{Poleski2008}, the finer time resolution adopted by \citet{Karczmarek2011} in their study could also make their analysis sensitive to erratic period changes that take place at much shorter timescales than probed by us, since we often binned together data spanning multiple years, in order to increase the statistical significance of our results and to more reliably analyze long-term trends.  

As a rule, the $O-C$ diagrams that we obtained for long-period Cepheids, i.e., Cepheids with periods longer than about 50~d, do not show the typical quadratic behavior that would have been expected in the case of linear period changes, but rather fluctuations, which sometimes can be very substantial, superimposed on a global long-term trend. Following \citet{Sterken2005}, we suggest that this behavior might be successfully described in terms of a combination of two terms, namely a quadratic one, reflecting the secular evolution of the physical properties of the star, and a cyclical, quasi-sinusoidal one, whose physical origin is still unclear. We speculate that the latter could in principle be caused by the presence of one or more companions, but the fact that a similar behavior is seen in all long-period Cepheids, and often with very high amplitudes in their corresponding $O-C$ diagrams, renders this an unlikely general explanation, as does the fact that the implied minimum masses of the companion would be in excess of $10^6 \, \rm{M}_{\sun}$ in all cases. This approximation for the masses was computed by using the binary mass function and the eye-approximate period (around 50~yr) and semi-major axis (about 2000~AU) of an edge-on view of the assumed binary orbit. 

\section*{Acknowledgements}

Support for this project is provided by ANID's Millennium Science Initiative through grant ICN12\textunderscore 12009, awarded to the Millennium Institute of Astrophysics (MAS); by ANID's Basal projects AFB-170002 and FB210003; and by FONDECYT grant \#1171273. The research leading to these results has received funding from the European Research Council (ERC) under the European Union’s Horizon 2020 research and innovation programme (grant agreement No 695099). CC, COH, and AVN acknowledge support from the National Agency for Research and Development (ANID), Scholarship Program Doctorado Nacional, grant numbers 2021\,–\,21212083, 2018\,–\,21180315, and 2020\,–\,21201226, respectively. The DASCH project at Harvard was partially supported by the NSF grants AST-0407380, AST-0909073, and AST-1313370. We are grateful for the efforts of the DASCH project for making the present study possible.

\section*{Data Availability}

The derived data generated in this article are available in its online supplementary material. This dataset was derived from sources in the public domain: the \href{http://www.astrouw.edu.pl/asas/?page=acvs}{ASAS Catalogue of Variable Stars}, the \href{http://dasch.rc.fas.harvard.edu/lightcurve.php}{DASCH Lightcurve Access} portal, the \href{https://gea.esac.esa.int/archive/}{\textit{Gaia} Archive}, the \href{http://ogle.astrouw.edu.pl/}{OGLE Project} and the \href{https://macho.nci.org.au/}{MACHO Data Services} .


\bibliographystyle{mnras}
\bibliography{main}

\begin{thebibliography}{}
\makeatletter
\relax
\def\mn@urlcharsother{\let\do\@makeother \do\$\do\&\do\#\do\^\do\_\do\%\do\~}
\def\mn@doi{\begingroup\mn@urlcharsother \@ifnextchar [ {\mn@doi@}
  {\mn@doi@[]}}
\def\mn@doi@[#1]#2{\def\@tempa{#1}\ifx\@tempa\@empty \href
  {http://dx.doi.org/#2} {doi:#2}\else \href {http://dx.doi.org/#2} {#1}\fi
  \endgroup}
\def\mn@eprint#1#2{\mn@eprint@#1:#2::\@nil}
\def\mn@eprint@arXiv#1{\href {http://arxiv.org/abs/#1} {{\tt arXiv:#1}}}
\def\mn@eprint@dblp#1{\href {http://dblp.uni-trier.de/rec/bibtex/#1.xml}
  {dblp:#1}}
\def\mn@eprint@#1:#2:#3:#4\@nil{\def\@tempa {#1}\def\@tempb {#2}\def\@tempc
  {#3}\ifx \@tempc \@empty \let \@tempc \@tempb \let \@tempb \@tempa \fi \ifx
  \@tempb \@empty \def\@tempb {arXiv}\fi \@ifundefined
  {mn@eprint@\@tempb}{\@tempb:\@tempc}{\expandafter \expandafter \csname
  mn@eprint@\@tempb\endcsname \expandafter{\@tempc}}}

\bibitem[\protect\citeauthoryear{{Alcock} et~al.,}{{Alcock}
  et~al.}{2000}]{Alcock2000}
{Alcock} C.,  et~al., 2000, \mn@doi [\apj] {10.1086/309512}, \href
  {https://ui.adsabs.harvard.edu/abs/2000ApJ...542..281A} {542, 281}

\bibitem[\protect\citeauthoryear{{Berdnikov}, {Samus}, {Antipin}, {Ezhkova},
  {Pastukhova}  \& {Turner}}{{Berdnikov} et~al.}{2004}]{Berdnikov2004}
{Berdnikov} L.~N.,  {Samus} N.~N.,  {Antipin} S.~V.,  {Ezhkova} O.~V.,
  {Pastukhova} E.~N.,   {Turner} D.~G.,  2004, \mn@doi [\pasp]
  {10.1086/420984}, \href
  {https://ui.adsabs.harvard.edu/abs/2004PASP..116..536B} {116, 536}

\bibitem[\protect\citeauthoryear{{Breuval} et~al.,}{{Breuval}
  et~al.}{2020}]{Breuval2020}
{Breuval} L.,  et~al., 2020, \mn@doi [\aap] {10.1051/0004-6361/202038633},
  \href {https://ui.adsabs.harvard.edu/abs/2020A&A...643A.115B} {643, A115}

\bibitem[\protect\citeauthoryear{{Catanzaro} et~al.,}{{Catanzaro}
  et~al.}{2020}]{Catanzaro2020}
{Catanzaro} G.,  et~al., 2020, \mn@doi [\aap] {10.1051/0004-6361/202038486},
  \href {https://ui.adsabs.harvard.edu/abs/2020A&A...639L...4C} {639, L4}

\bibitem[\protect\citeauthoryear{{Catelan} \& {Smith}}{{Catelan} \&
  {Smith}}{2015}]{Catelan2015}
{Catelan} M.,  {Smith} H.~A.,  2015, {Pulsating Stars}.
Wiley-VCH

\bibitem[\protect\citeauthoryear{{Evans} et~al.,}{{Evans}
  et~al.}{2018}]{GaiaDR2-phot2018}
{Evans} D.~W.,  et~al., 2018, \mn@doi [\aap] {10.1051/0004-6361/201832756},
  \href {https://ui.adsabs.harvard.edu/abs/2018A&A...616A...4E} {616, A4}

\bibitem[\protect\citeauthoryear{{Foreman-Mackey}, {Hogg}, {Lang}  \&
  {Goodman}}{{Foreman-Mackey} et~al.}{2013}]{ForemanMackey2013}
{Foreman-Mackey} D.,  {Hogg} D.~W.,  {Lang} D.,   {Goodman} J.,  2013, \mn@doi
  [\pasp] {10.1086/670067}, \href
  {https://ui.adsabs.harvard.edu/abs/2013PASP..125..306F} {125, 306}

\bibitem[\protect\citeauthoryear{{Freedman} et~al.,}{{Freedman}
  et~al.}{2001}]{Freedman-2001}
{Freedman} W.~L.,  et~al., 2001, \mn@doi [\apj] {10.1086/320638}, \href
  {https://ui.adsabs.harvard.edu/abs/2001ApJ...553...47F} {553, 47}

\bibitem[\protect\citeauthoryear{{Gaia Collaboration} et~al.,}{{Gaia
  Collaboration} et~al.}{2018}]{GaiaCollaboration2018}
{Gaia Collaboration} et~al., 2018, \mn@doi [\aap]
  {10.1051/0004-6361/201833051}, \href
  {https://ui.adsabs.harvard.edu/abs/2018A&A...616A...1G} {616, A1}

\bibitem[\protect\citeauthoryear{{Grindlay}, {Tang}, {Los}  \&
  {Servillat}}{{Grindlay} et~al.}{2012}]{Grindlay2012}
{Grindlay} J.,  {Tang} S.,  {Los} E.,   {Servillat} M.,  2012, in {Griffin} E.,
   {Hanisch} R.,   {Seaman} R.,  eds,  IAU Symposium Vol. 285, New Horizons in
  Time Domain Astronomy. pp 29--34 (\mn@eprint {arXiv} {1211.1051}),
  \mn@doi{10.1017/S1743921312000166}

\bibitem[\protect\citeauthoryear{{Hajdu}, {Catelan}, {Jurcsik}, {Dekany},
  {Drake}  \& {Marquette}}{{Hajdu} et~al.}{2015}]{Hajdu2015}
{Hajdu} G.,  {Catelan} M.,  {Jurcsik} J.,  {Dekany} I.,  {Drake} A.~J.,
  {Marquette} J.~B.,  2015, \mn@doi [\mnras] {10.1093/mnrasl/slv024}, \href
  {https://ui.adsabs.harvard.edu/abs/2015MNRAS.449L.113H} {449, L113}

\bibitem[\protect\citeauthoryear{{Hajdu} et~al.,}{{Hajdu}
  et~al.}{2021}]{Hajdu-2021}
{Hajdu} G.,  et~al., 2021, \mn@doi [\apj] {10.3847/1538-4357/abff4b}, \href
  {https://ui.adsabs.harvard.edu/abs/2021ApJ...915...50H} {915, 50}

\bibitem[\protect\citeauthoryear{{Henden}, {Smith}, {Levine}  \&
  {Terrell}}{{Henden} et~al.}{2012}]{Henden2012}
{Henden} A.~A.,  {Smith} T.~C.,  {Levine} S.~E.,   {Terrell} D.,  2012, in
  American Astronomical Society Meeting Abstracts \#220. p. 133.06

\bibitem[\protect\citeauthoryear{{Hertzsprung}}{{Hertzsprung}}{1919}]{Her1919}
{Hertzsprung} E.,  1919, \mn@doi [Astronomische Nachrichten]
  {10.1002/asna.19202100202}, \href
  {https://ui.adsabs.harvard.edu/abs/1919AN....210...17H} {210, 17}

\bibitem[\protect\citeauthoryear{{Jurcsik} et~al.,}{{Jurcsik}
  et~al.}{2012}]{Jurcsik-2012}
{Jurcsik} J.,  et~al., 2012, \mn@doi [\mnras]
  {10.1111/j.1365-2966.2011.19868.x}, \href
  {https://ui.adsabs.harvard.edu/abs/2012MNRAS.419.2173J} {419, 2173}

\bibitem[\protect\citeauthoryear{{Karczmarek}, {Dziembowski}, {Lenz},
  {Pietrukowicz}  \& {Pojma{\'n}ski}}{{Karczmarek}
  et~al.}{2011}]{Karczmarek2011}
{Karczmarek} P.,  {Dziembowski} W.~A.,  {Lenz} P.,  {Pietrukowicz} P.,
  {Pojma{\'n}ski} G.,  2011, \actaa, \href
  {https://ui.adsabs.harvard.edu/abs/2011AcA....61..303K} {61, 303}

\bibitem[\protect\citeauthoryear{{Kovtyukh} et~al.,}{{Kovtyukh}
  et~al.}{2019}]{Kov2019}
{Kovtyukh} V.,  et~al., 2019, \mn@doi [\mnras] {10.1093/mnras/stz1872}, \href
  {https://ui.adsabs.harvard.edu/abs/2019MNRAS.488.3211K} {488, 3211}

\bibitem[\protect\citeauthoryear{{Lasker} et~al.,}{{Lasker}
  et~al.}{2008}]{Lasker2008}
{Lasker} B.~M.,  et~al., 2008, \mn@doi [\aj] {10.1088/0004-6256/136/2/735},
  \href {https://ui.adsabs.harvard.edu/abs/2008AJ....136..735L} {136, 735}

\bibitem[\protect\citeauthoryear{{Le Borgne} et~al.,}{{Le Borgne}
  et~al.}{2007}]{LeBorgne2007}
{Le Borgne} J.~F.,  et~al., 2007, \mn@doi [\aap] {10.1051/0004-6361:20077957},
  \href {https://ui.adsabs.harvard.edu/abs/2007A&A...476..307L} {476, 307}

\bibitem[\protect\citeauthoryear{{Leavitt} \& {Pickering}}{{Leavitt} \&
  {Pickering}}{1912}]{Leavitt1912}
{Leavitt} H.~S.,  {Pickering} E.~C.,  1912, Harvard College Observatory
  Circular, \href {https://ui.adsabs.harvard.edu/abs/1912HarCi.173....1L} {173,
  1}

\bibitem[\protect\citeauthoryear{{Luck}, {Kovtyukh}  \& {Andrievsky}}{{Luck}
  et~al.}{2001}]{Luck2001}
{Luck} R.~E.,  {Kovtyukh} V.~V.,   {Andrievsky} S.~M.,  2001, \mn@doi [\aap]
  {10.1051/0004-6361:20010615}, \href
  {https://ui.adsabs.harvard.edu/abs/2001A&A...373..589L} {373, 589}

\bibitem[\protect\citeauthoryear{{Neilson} \& {Ignace}}{{Neilson} \&
  {Ignace}}{2014}]{Neilson2014}
{Neilson} H.~R.,  {Ignace} R.,  2014, \mn@doi [\aap]
  {10.1051/0004-6361/201423444}, \href
  {https://ui.adsabs.harvard.edu/abs/2014A&A...563L...4N} {563, L4}

\bibitem[\protect\citeauthoryear{{Paxton}, {Bildsten}, {Dotter}, {Herwig},
  {Lesaffre}  \& {Timmes}}{{Paxton} et~al.}{2011}]{Paxton-2011}
{Paxton} B.,  {Bildsten} L.,  {Dotter} A.,  {Herwig} F.,  {Lesaffre} P.,
  {Timmes} F.,  2011, \mn@doi [\apjs] {10.1088/0067-0049/192/1/3}, \href
  {https://ui.adsabs.harvard.edu/abs/2011ApJS..192....3P} {192, 3}

\bibitem[\protect\citeauthoryear{{Paxton} et~al.,}{{Paxton}
  et~al.}{2019}]{Paxton2019}
{Paxton} B.,  et~al., 2019, \mn@doi [\apjs] {10.3847/1538-4365/ab2241}, \href
  {https://ui.adsabs.harvard.edu/abs/2019ApJS..243...10P} {243, 10}

\bibitem[\protect\citeauthoryear{{Payne-Gaposchkin}}{{Payne-Gaposchkin}}{1971}]{PayneGaposchkin1971}
{Payne-Gaposchkin} C.~H.,  1971, Smithsonian Contributions to Astrophysics,
  \href {https://ui.adsabs.harvard.edu/abs/1971SCoA...13.....P} {13}

\bibitem[\protect\citeauthoryear{{Pietrukowicz}}{{Pietrukowicz}}{2001}]{Pietrukowicz2001}
{Pietrukowicz} P.,  2001, \actaa, \href
  {https://ui.adsabs.harvard.edu/abs/2001AcA....51..247P} {51, 247}

\bibitem[\protect\citeauthoryear{{Pojmanski}}{{Pojmanski}}{2002}]{Pojmanski2002}
{Pojmanski} G.,  2002, \actaa, \href
  {https://ui.adsabs.harvard.edu/abs/2002AcA....52..397P} {52, 397}

\bibitem[\protect\citeauthoryear{{Poleski}}{{Poleski}}{2008}]{Poleski2008}
{Poleski} R.,  2008, \actaa, \href
  {https://ui.adsabs.harvard.edu/abs/2008AcA....58..313P} {58, 313}

\bibitem[\protect\citeauthoryear{{Riess}, {Casertano}, {Yuan}, {Bowers},
  {Macri}, {Zinn}  \& {Scolnic}}{{Riess} et~al.}{2021}]{Riess2021}
{Riess} A.~G.,  {Casertano} S.,  {Yuan} W.,  {Bowers} J.~B.,  {Macri} L.,
  {Zinn} J.~C.,   {Scolnic} D.,  2021, \mn@doi [\apjl]
  {10.3847/2041-8213/abdbaf}, \href
  {https://ui.adsabs.harvard.edu/abs/2021ApJ...908L...6R} {908, L6}

\bibitem[\protect\citeauthoryear{{Ripepi} et~al.,}{{Ripepi}
  et~al.}{2021}]{Ripepi2021}
{Ripepi} V.,  et~al., 2021, \mn@doi [\aap] {10.1051/0004-6361/202040123}, \href
  {https://ui.adsabs.harvard.edu/abs/2021A&A...647A.111R} {647, A111}

\bibitem[\protect\citeauthoryear{{Ritter}}{{Ritter}}{1879}]{Ritter1879}
{Ritter} A.,  1879, \mn@doi [Annalen der Physik] {10.1002/andp.18792440910},
  \href {https://ui.adsabs.harvard.edu/abs/1879AnP...244..157R} {244, 157}

\bibitem[\protect\citeauthoryear{{Seidelmann} \& {Fukushima}}{{Seidelmann} \&
  {Fukushima}}{1992}]{Seidelmann1992}
{Seidelmann} P.~K.,  {Fukushima} T.,  1992, \aap, \href
  {https://ui.adsabs.harvard.edu/abs/1992A&A...265..833S} {265, 833}

\bibitem[\protect\citeauthoryear{{Silva Aguirre}, {Catelan}, {Weiss}  \&
  {Valcarce}}{{Silva Aguirre} et~al.}{2008}]{Silva-Aguirre-2008}
{Silva Aguirre} V.,  {Catelan} M.,  {Weiss} A.,   {Valcarce} A.~A.~R.,  2008,
  \mn@doi [\aap] {10.1051/0004-6361:200810047}, \href
  {https://ui.adsabs.harvard.edu/abs/2008A&A...489.1201S} {489, 1201}

\bibitem[\protect\citeauthoryear{{Soszynski} et~al.,}{{Soszynski}
  et~al.}{2008}]{Soszynski2008}
{Soszynski} I.,  et~al., 2008, \actaa, \href
  {https://ui.adsabs.harvard.edu/abs/2008AcA....58..163S} {58, 163}

\bibitem[\protect\citeauthoryear{{Soszy{\'n}ski} et~al.,}{{Soszy{\'n}ski}
  et~al.}{2015}]{Soszynski2015}
{Soszy{\'n}ski} I.,  et~al., 2015, \actaa, \href
  {https://ui.adsabs.harvard.edu/abs/2015AcA....65..297S} {65, 297}

\bibitem[\protect\citeauthoryear{{Sterken}}{{Sterken}}{2005}]{Sterken2005}
{Sterken} C.,  2005, in {Sterken} C.,  ed.,  Astronomical Society of the
  Pacific Conference Series Vol. 335, The Light-Time Effect in Astrophysics:
  Causes and cures of the O-C diagram. p.~3

\bibitem[\protect\citeauthoryear{{Szabados}}{{Szabados}}{1983}]{Szabados1983}
{Szabados} L.,  1983, \mn@doi [\apss] {10.1007/BF00661952}, \href
  {https://ui.adsabs.harvard.edu/abs/1983Ap&SS..96..185S} {96, 185}

\bibitem[\protect\citeauthoryear{{Tang}, {Grindlay}, {Los}  \&
  {Servillat}}{{Tang} et~al.}{2013}]{Tang2013}
{Tang} S.,  {Grindlay} J.,  {Los} E.,   {Servillat} M.,  2013, \mn@doi [\pasp]
  {10.1086/671760}, \href
  {https://ui.adsabs.harvard.edu/abs/2013PASP..125..857T} {125, 857}

\bibitem[\protect\citeauthoryear{{Taylor}}{{Taylor}}{2005}]{Taylor2005}
{Taylor} M.~B.,  2005, in {Shopbell} P.,  {Britton} M.,   {Ebert} R.,  eds,
  Astronomical Society of the Pacific Conference Series Vol. 347, Astronomical
  Data Analysis Software and Systems XIV. p.~29

\bibitem[\protect\citeauthoryear{{Turner}, {Abdel-Sabour Abdel-Latif}  \&
  {Berdnikov}}{{Turner} et~al.}{2006}]{Turner2006}
{Turner} D.~G.,  {Abdel-Sabour Abdel-Latif} M.,   {Berdnikov} L.~N.,  2006,
  \mn@doi [\pasp] {10.1086/499501}, \href
  {https://ui.adsabs.harvard.edu/abs/2006PASP..118..410T} {118, 410}

\bibitem[\protect\citeauthoryear{{Udalski}, {Szyma{\'n}ski}  \&
  {Szyma{\'n}ski}}{{Udalski} et~al.}{2015}]{Udalski2015}
{Udalski} A.,  {Szyma{\'n}ski} M.~K.,   {Szyma{\'n}ski} G.,  2015, \actaa,
  \href {https://ui.adsabs.harvard.edu/abs/2015AcA....65....1U} {65, 1}

\bibitem[\protect\citeauthoryear{{Ulaczyk} et~al.,}{{Ulaczyk}
  et~al.}{2012}]{Ulaczyk2012}
{Ulaczyk} K.,  et~al., 2012, \actaa, \href
  {https://ui.adsabs.harvard.edu/abs/2012AcA....62..247U} {62, 247}

\bibitem[\protect\citeauthoryear{{Zhou}}{{Zhou}}{1999}]{Zhou1999}
{Zhou} A.,  1999, Publications of the Beijing Astronomical Observatory, \href
  {https://ui.adsabs.harvard.edu/abs/1999PBeiO..33...17Z} {33, 17}

\makeatother
\end{thebibliography}

\bsp	
\label{lastpage}
\end{document}